\documentclass[aps,prb,preprint,unsortedaddress,showpacs,citeautoscript]{revtex4}
\usepackage{amssymb}

\usepackage{amsmath}
\usepackage{graphicx}
\newcommand{\rb}[1]{\raisebox{1.5ex}[-1.5ex]{#1}}

\begin{document}

\title{Thermodynamics of Two\,-\,Band Superconductors: The Case of MgB$_{2}$}
\author{Oleg V. Dolgov}\email{O.Dolgov@fkf.mpg.de}
\author{Reinhard K. Kremer}
\affiliation{Max-Planck-Institut f{\"u}r Festk{\"o}rperforschung, Heisenbergstr. 1, D-70569
Stuttgart, Germany}

\author{Jens Kortus}
\affiliation{Max-Planck-Institut f{\"u}r Festk{\"o}rperforschung, Heisenbergstr. 1, D-70569
Stuttgart, Germany}
\affiliation{Institut de Physique et de
Chimie des Mat\'eriaux de Strasbourg, 23 Rue du Loess, F-67034
Strasbourg Cedex 2, France}

\author{Alexander A. Golubov}
\affiliation{Faculty of Science and Technology, University of Twente, 7500 AE Enschede,
The Netherlands}
\author{Sergei V. Shulga}
\affiliation{Institut f\"{u}r Festk\"{o}rper und Werkstofforschung Dresden e.V., Postfach
270016, D-01171 Dresden, Germany }
\date{\today }

\begin{abstract}
Thermodynamic properties of the multiband superconductor MgB$_{2}$
have often been described using a
simple sum of the  standard BCS expressions corresponding to $\sigma$- and
$\pi$-bands. Although, it is \textit{a priori} not clear if this
approach is working always adequately, in particular in cases of
strong interband scattering. Here we compare the often used
approach of a sum of two independent bands using BCS-like
$\alpha$-model expressions for the specific heat, entropy and free
energy to the solution of the full Eliashberg equations. The
superconducting energy gaps, the free energy, the entropy and the
heat capacity for varying interband scattering rates are
calculated within the framework of two-band Eliashberg theory.
We obtain good agreement between the phenomenological two-band
$\alpha$-model  with the Eliashberg results, which delivers for
the first time the theoretical verification to use the
$\alpha$-model as a useful tool for a reliable analysis of heat
capacity data. For the thermodynamic potential and the entropy we
demonstrate that only the sum over the contributions of the two
bands has physical meaning.
\end{abstract}

\pacs{74.25Bt, 74.70.Ad, 74.62.Dh}
 \maketitle

\section{Introduction}

Apart from the high transition temperature of 40\thinspace K
\cite{Nagamatsu}, two-band superconductivity was the other unexpected phenomenon
in MgB$_{2}$ which attracts increasing attention. In fact, at
present it appears that MgB$_{2}$ is the only superconductor with
substantiated theoretical and experimental evidence for two-band
superconductivity.

Historically, two-band superconductivity has already been
investigated theoretically shortly after the formulation of BCS
theory. Suhl, Matthias and Walker \cite{Suhl} suggested a model
for transition metals considering overlapping $s$- and
$d$-bands. At the same time, Moskalenko proposed an extension of
BCS theory for multiple bands \cite{Moskalenko}.
A review of theoretical treatment of the critical temperature
$T_{c}$ of multiband superconductors may be found in
Ref.\onlinecite{Allen}.

In the early sixties there have been experimental claims for the
observation of two-band superconductivity in some transition
metals like e.g.\ V, Nb and Ta \cite{Shen,Radebough} and later  in
oxygen depleted SrTiO$_3$ \cite{Bednorz}.

Until now, MgB$_{2}$ appears to be first system for which
multi-band superconductivity has independently been evidenced by
several experimental techniques like, for example,
heat capacity, tunneling spectroscopy, Raman
spectroscopy, penetration depth measurements, ARPRES, and the
analysis of the critical fields \cite{PhysicaC}. The theoretical
justification for two-band superconductivity in MgB$_{2}$ has been given
from electronic structure calculations \cite{Kortus,Pickett}.
These find that the Fermi surface contains two quasi-cylindrical sheets
corresponding to nearly two-dimensional $\sigma$-bands. A three dimensional
network of the Fermi surface is attributed to the $\pi$-bands. It
has been demonstrated that the optical bond stretching
$E_{\mathrm{2g}}$ phonons couple strongly to the holes at the top
of $\sigma$-bands while the three-dimensional $\pi$-electrons
couple only weakly to the phonons.
The different coupling strengths of the $\sigma$- and $\pi$-bands
lead to superconducting gaps different in character and size
\cite{Liu,Kong,Bohnen,Choi,Kunc}. Using linear response theory it is
possible to calculate the electron-phonon coupling
(Eliashberg functions) from first principles. The
superconducting gaps obtained from Eliashberg theory
are in very good agreement with the experiments \cite{Golubov,Choi}.

Interband scattering from  impurities  will complicate this
picture because interband scattering leads to a decrease of
$T_{\mathrm{c}}$ and finally to a single order parameter
\cite{Schopohl,GolMaz,Golubov,Brinkman}. Interband scattering is
weak in MgB$_{2}$ \cite{mazinimp}, but this is not necessarily the
case in samples in which Mg has been replaced by Al or B by C
(Ref.\onlinecite
{Agrestini,Bouquet,Junod,Xiang,Li,Margadonna,Papavass,Bianconi,Pena,Postorino,
Castro,Putti,Ribeiro,Schmidt,Papagelis,Holanova,Gonelli1}) or
which have been exposed to neutron irradiation \cite{Wangirra}.
Such samples exhibit considerably reduced $T_{\mathrm{c}}$, while
the two gaps persist even at very low critical temperatures.
Recently it was shown that the $T_{\mathrm{c}}$ reduction in
MgB$_{2}$ due to Al or C doping can be explained mainly as due to
a simple effect of band filling \cite{PRL2005,Ummarino}. A similar
observation has been made using a phenomenological weak-coupling
approach\cite{Bussmann}. Further, the doping independent $\pi$-gap
in C-doped MgB$_{2}$ can be understood as due to a compensation of
band filling and interband scattering effects.

Thermodynamic properties of anisotropic superconductors in the
weak coupling regime were extensively studied in the past. In the
case of weak anisotropy the BCS model was extended by Pokrovsky
\cite{Pokrov}. It was shown that the specific heat jump at $T_{c}$
is reduced as compared to the isotropic case. For two-band weakly
coupled superconductors the specific heat was calculated
by several authors \cite%
{Moscal91,Palistrant92,SodaWada,Geilikman,ShenSonPhil,Kresin73,%
Chi,Mishonov,Ramunni,Nakai,Kristof,Watanabe}
(for a recent review see also Ref. \onlinecite{Palistrant}).
The main prediction is that at $T_{c}$ the relative jump in
the electronic specific heat, $(C_{SC}-C_{N})/C_{N}$, is
reduced as compared to the universal BCS value of 1.43.
On the other hand, for an isotropic strongly coupled superconductor
the relative specific heat jump is larger than 1.43
(see e.g.\ the review in Ref. \onlinecite{Carbotte}).
The combined effect of strong coupling and multiband anisotropy
on the specific heat was studied earlier by the present authors
\cite{Golubov}, where the results of the first principles
calculations of the electron-phonon interaction in MgB$_{2}$ were
used but the effect of interband impurity scattering was not
considered. Recently strong-coupling corrections were taken into
account in the so-called two-square-well approximation (separable model)
\cite{Zehet,Mitr,Nicol}, where the effect of interband scattering on
some thermodynamic functions was studied \cite{Mitr,Nicol}.

In the present work we formulate a generalized description of
the thermodynamics of multiband superconductors taking into
account impurity scattering (magnetic and nonmagnetic) in the
framework of two-band Eliashberg theory. The results are
applied to MgB$_{2}$ using the first principles band-structure
results for the electronic spectra and electron-phonon interaction
\cite{Kong} by extending our preceding approach \cite{Golubov}. The
superconducting energy gaps, the free energy, the entropy and the
heat capacity for varying nonmagnetic interband scattering rates
are calculated within the framework of two-band Eliashberg
theory. It will be shown that the expression for the thermodynamic
potential on the extremal trajectory corresponding to
solutions of the Eliashberg equations has the form of the sum of
contributions of $\sigma$- and $\pi$-bands, but that only the
total thermodynamic potential (the sum of both contributions) has
physical meaning.

In a second step, we perform a  comparison of the phenomenological
two-band $\alpha$-model with the Eliashberg results and apply a
fit program developed for the $\alpha$-model to extract the gaps
and the Sommerfeld constants from the Eliashberg results. Good
agreement of the two band $\alpha$-model with the Eliashberg data
is found for the temperature dependence of total heat capacity,
the entropy and the free energy and the gaps. There are, however,
distinct deviations in the partial contributions to the individual
quantities and the Sommerfeld constants obtained from the fits. We
conclude that the phenomenological $\alpha$-model approach can be
taken as a handy tool to analyze e. g. experimental heat capacity
data  and the gaps to a satisfying accuracy, however that care
must be taken for other quantities.

The paper is organized as follows: In Section II the introduction
to the formalism and the method of solution is given, in Section
III numerical results for the densities of states (DOS) and various
thermodynamic quantities as a function of interband impurity
scattering rate are discussed, in Section IV the comparison of the
two-band $\alpha$-model with the Eliashberg results is performed.
In the Appendix a general expression for the thermodynamic potential
of a multiband superconductor with nonmagnetic impurities is
derived.

\section{Free energy and Eliashberg Equations}

A general expression for the difference of free energies
$\Delta`\Omega =\Omega _{N}-\Omega _{S}$ in the normal (N) and
superconducting (S) state for a system with electron-phonon
interaction and multiple bands can be obtained in two ways: one
has been derived by a straightforward integration
over the electron-phonon interaction constants by Golubov \textit{et al.}
\cite{Golubov}. The derivation of the expression for the thermodynamic
potential for the case of nonmagnetic as well magnetic impurities is
presented in Appendix A.
In terms of Matsubara frequencies the $\Omega$-potential can be written as
\begin{eqnarray*}
\Omega &=&\Omega _{e}^{(0)}+\Omega _{\mathrm{ph}}^{(0)}-2\pi
\sum_{i}N_{i}(0)T\sum\limits_{n}\frac{\omega _{n}^{2}(Z_{in}^{2}-1)+\Phi
_{in}^{2}}{|\omega _{n}|+\sqrt{\omega _{n}^{2}Z_{in}^{2}+\Phi _{in}^{2}}}+ \\
&&+\pi \sum_{i}N_{i}(0)T\sum\limits_{n}\frac{\omega
_{n}^{2}Z_{in}(Z_{in}-1)+\Phi _{in}^{2}}{\sqrt{\omega
_{n}^{2}Z_{in}^{2}+\Phi _{in}^{2}}},
\end{eqnarray*}%
where $\Omega _{e}^{(0)}$ is the $\Omega $-potential of the noninteracting
electrons, and $\Omega _{\mathrm{ph}}^{(0)}$ is the $\Omega $-potential of
the noninteracting phonons.

For the difference of the $\Omega$-potentials in the normal and the
superconducting state one obtains
\begin{eqnarray}
\Delta \Omega &=&\Omega _{N}-\Omega _{S}=-\pi
\sum_{i}N_{i}(0)T\sum\limits_{n=-\omega _{c}}^{\omega _{c}}
\label{eq:wefree} \\
&&\left\{ |\omega _{n}|(Z_{in}^{N}-1)-\frac{2\omega _{n}^{2}[\left(
Z_{in}^{S}\right) ^{2}-1]+2\Phi _{in}^{2}}{|\omega _{n}|+\sqrt{\omega
_{n}^{2}\left( Z_{in}^{S}\right) ^{2}+\Phi _{in}^{2}}}+\frac{\omega
_{n}^{2}Z_{in}^{S}(Z_{in}^{S}-1)+\Phi _{in}^{2}}{\sqrt{\omega _{n}^{2}\left(
Z_{in}^{S}\right) ^{2}+\Phi _{in}^{2}}}\right\} .  \notag
\end{eqnarray}
$Z$ is the renormalization factor (which is unity in the weak
coupling limit)
and $\Phi $ is the order parameter which is connected to the energy gap via
$\Delta_i(\omega_n)=\Phi_{i}(\omega_n)/Z_{i}(\omega_n) =\Phi_{in}/Z_{in}$.
$Z^N$ and $Z^S$ correspond to the normal state ($\Delta$ = 0) and the
superconducting state, respectively.
The summations in Eq.~\ref{eq:wefree} are carried out over the
fermionic Matsubara (temperature) frequencies $\omega _{n}=\pi
T(2n-1)$ as well as over the band index $i=\sigma ,\pi$.
$N_{i}(0)$ are the partial electronic DOS's for the $\sigma$- and $\pi$-bands
at the Fermi level.

Another way to express the free energy was suggested by Carbotte
\cite{Carbotte}, who proposed that it is possible to find a functional, which
minimization with respect to $Z$ and $\Phi$ gives the Eliashberg relations
for superconductors with strong electron-boson interaction.
For a multiband system the corresponding functional is given by
\begin{equation*}
\Delta \mathcal{F}=2\pi T\sum_{i,n}N_{i}(0)\omega _{n}\left( \frac{%
Z_{in}\omega _{n}}{\sqrt{\left( Z_{in}\omega _{n}\right) ^{2}+\Phi _{in}^{2}}%
}-\text{sign}\omega _{n}\right)
\end{equation*}%
\begin{equation*}
+\pi ^{2}T^{2}\sum_{n,m}\sum_{i,j}N_{i}(0)\left[ \frac{Z_{in}\omega _{n}}{%
\sqrt{\left( Z_{in}\omega _{n}\right) ^{2}+\Phi _{in}^{2}}}\frac{%
Z_{m,j}\omega _{m}}{\sqrt{\left( Z_{jm}\omega _{m}\right) ^{2}+\Phi _{jm}^{2}%
}}-\text{sign}(\omega _{n}\omega _{m})\right] \tilde{\lambda}%
_{ij}^{+}(\omega _{n}-\omega _{m})
\end{equation*}%
\begin{equation}
+\pi ^{2}T^{2}\sum_{n,m}^{|\omega _{c}|}\sum_{i,j}N_{i}(0)\frac{\Phi _{in}}{%
\sqrt{\left( Z_{in}\omega _{n}\right) ^{2}+\Phi _{in}^{2}}}\frac{\Phi _{jm}}{%
\sqrt{\left( Z_{jm}\omega _{m}\right) ^{2}+\Phi _{jm}^{2}}}\left[ \tilde{%
\lambda}_{ij}^{-}(\omega _{n}-\omega _{m})-\mu _{ij}^{\ast }(\omega _{c})%
\right],  \label{frcarb}
\end{equation}
where

\begin{equation*}
\tilde{\lambda}_{ij}^{\pm }(\omega _{n}-\omega _{m})=\int_{0}^{\infty }\frac{%
d\omega ^{2}\alpha _{ij}^{2}(\omega )F_{ij}(\omega )}{(\omega _{n}-\omega
_{m})^{2}+\omega ^{2}}+\left( \Gamma _{ij}\pm \Gamma _{ij}^{m}\right) \delta
_{n,m}/2\pi T
\end{equation*}
represents the electron-phonon interaction together with
nonmagnetic $\Gamma _{ij}$ and magnetic $\Gamma _{ij}^{m}$
impurity scattering terms. The Coulomb pseudopotential
$\mu_{ij}^{\ast }(\omega _{c})$ is determined at a frequency
$\omega_{c}$ which has to be chosen much larger than the maximal phonon
frequency. Minimization of Eq. \ref{frcarb} provides the
Eliashberg equations on the imaginary (Matsubara) axis

\begin{eqnarray}
Z_{ni}\,\omega _{n} &=&\omega _{n}+\pi T\sum_{m,j}\tilde{\lambda}%
_{ij}^{+}(\omega _{n}-\omega _{m})\frac{Z_{jm}\omega _{m}}{\sqrt{\left(
Z_{jm}\omega _{n}\right) ^{2}+\Phi _{jm}^{2}}},  \label{matsEE} \\
\Phi _{ni} &=&\pi T\sum_{m,j}^{\left| \omega _{m}\right| \leq |\omega _{c}|}
\left[ \tilde{\lambda}_{ij}^{-}(\omega _{n}-\omega _{m})-\mu _{ij}^{\ast
}(\omega _{c})\right] \frac{\Phi _{jm}}{\sqrt{\left( Z_{jm}\omega
_{m}\right) ^{2}+\Phi _{jm}^{2}}}.  \notag
\end{eqnarray}

Both functionals $\Delta \Omega $ and $\Delta \mathcal{F}$ give
the same result on the extremal trajectory which corresponds to
solutions of the Eliashberg equations Eq.(\ref{matsEE}). In
the following we will use $\Delta \Omega $ for the calculations of
thermodynamic quantities.

For the calculations e.g., of the densities of states
$N_{i}(\omega )$ in the superconducting state, we need to know the renormalization
factors $Z_{i}$ and the order parameters $\Phi_{i}$ along the real
frequency axis. The corresponding analytical continuation of
Eq.\ (\ref{matsEE}) substituting $i\omega _{n}\Longrightarrow \omega +i\delta$ gives
\begin{eqnarray}
Z_{i}(\omega )\omega &=&\omega -\int_{-\infty }^{\infty }d\omega ^{\prime
}\sum_{j}\int_{0}^{\infty }d\nu \alpha _{ij}^{2}(\nu )F_{ij}(\nu )I(\omega
+i\delta ,\nu ,\omega ^{\prime })\mbox{Re}\frac{Z_{j}(\omega ^{\prime
})\omega ^{\prime }}{\sqrt{(Z_{j}(\omega ^{\prime })\omega ^{\prime
})^{2}-\Phi _{j}^{2}(\omega ^{\prime })}}+  \notag \\
&&+i\sum_{j}\left( \Gamma _{ij}+\Gamma _{ij}^{m}\right) \frac{Z_{j}(\omega
^{\prime })\omega ^{\prime }}{\sqrt{(Z_{j}(\omega ^{\prime })\omega ^{\prime
})^{2}-\Phi _{j}^{2}(\omega ^{\prime })}},  \label{reEEa}
\end{eqnarray}

\begin{eqnarray}
\Phi _{i}(\omega ) &=&-\int_{-\infty }^{\infty }d\omega ^{\prime
}\sum_{j}\int_{0}^{\infty }d\nu \alpha _{ij}^{2}(\nu )F_{ij}(\nu )I(\omega
+i\delta ,\nu ,\omega ^{\prime })\mbox{Re}\frac{\Phi _{j}(\omega ^{\prime })%
}{\sqrt{(Z_{j}(\omega ^{\prime })\omega ^{\prime })^{2}-\Phi _{j}^{2}(\omega
^{\prime })}}-  \notag \\
&&-1/2\sum_{j}\mu _{ij}^{\ast }(\omega _{c})\int_{-\omega _{c}}^{\omega
_{c}}d\omega ^{\prime }\tanh \left( \frac{\omega ^{\prime }}{2T}\right) %
\mbox{Re}\frac{\Phi _{j}(\omega ^{\prime })}{\sqrt{(Z_{j}(\omega ^{\prime
})\omega ^{\prime })^{2}-\Phi _{j}^{2}(\omega ^{\prime })}}+  \notag \\
&&+i\sum_{j}\left( \Gamma _{ij}-\Gamma _{ij}^{m}\right) \frac{\Phi
_{j}(\omega ^{\prime })}{\sqrt{(Z_{j}(\omega ^{\prime })\omega ^{\prime
})^{2}-\Phi _{j}^{2}(\omega ^{\prime })}},  \label{reEEb}
\end{eqnarray}%
where%
\begin{equation*}
I(\omega +i\delta ,\nu ,\omega ^{\prime })=\frac{n(\nu )+1-f(\omega ^{\prime
})}{\omega +i\delta -\nu -\omega ^{\prime }}+\frac{n(\nu )+f(\omega ^{\prime
})}{\omega +i\delta +\nu -\omega ^{\prime }},
\end{equation*}%
and $n(\nu )$ and $f(\omega ^{\prime })$ are Bose and Fermi
distribution functions, respectively. Please note that the
nondiagonal elements of all functions $\alpha _{ij}^{2}(\nu
)F_{ij}(\nu ),$ $\mu _{ij}^{\ast }(\omega _{c})$, $\Gamma
_{ij}$,and $\Gamma _{ij}^{m}$ have to satisfy the
requirement of the detailed balance principle
\begin{equation*}
N_{\sigma }(0)\Gamma_{\sigma \pi }=N_{\pi }(0)\Gamma_{\pi \sigma },
\end{equation*}%
where $N_{\sigma }(0)$ and $N_{\pi}(0)$ are the normal state bare
electronic DOS's in the $\sigma$- and the $\pi$-bands,
respectively.

As has been shown in Ref.~\onlinecite{Mitr} and
\onlinecite{Nicol}, for a separable interaction $\lambda
_{ij}(\omega _{n}-\omega _{m})=\lambda _{ij}\theta (\Omega -\left|
\omega _{n}\right| )\theta (\Omega -\left| \omega _{m}\right| )$
the standard weak coupling expressions can be obtained which
correspond to the two-band BCS results.

In this paper we will use the following representation of the Eliashberg
equations which is better suited for numerical solution by iterations
\cite{Shulga}
\begin{eqnarray}
\mbox{Im}\,\Phi _{i}(\omega ) &=&\sum_{j}\frac{\left( \Gamma
_{ij}-\Gamma
_{ij}^{m}\right) }{2}\frac{\Phi _{j}(\omega )}{\sqrt{\tilde{\omega}%
_{j}^{2}(\omega )-\Phi _{j}^{2}(\omega )}}+\frac{\pi }{2}\sum_{j}\int
dy\alpha _{ij}^{2}F_{ij}(\omega -y)  \notag \\
&&\left[ \coth {\left( \frac{\omega -y}{2T}\right) }-\tanh {\left( \frac{y}{%
2T}\right) }\right] \mbox{Re}\frac{\Phi _{j}(y)}{\sqrt{\tilde{\omega}%
_{j}^{2}(y)-\Phi _{j}^{2}(y)}}  \label{eq1} \\
\mbox{Im}\,\tilde{\omega}_{i}(\omega ) &=&\sum_{j}\frac{\left(
\Gamma
_{ij}+\Gamma _{ij}^{m}\right) }{2}\frac{\tilde{\omega}_{j}(\omega )}{\sqrt{%
\tilde{\omega}_{j}^{2}(\omega )-\Phi j(\omega )}}+\frac{\pi }{2}\sum_{j}\int
dy\alpha _{ij}^{2}F_{ij}(\omega -y),  \notag \\
&&\left[ \coth {\left( \frac{\omega -y}{2T}\right) }-\tanh {\left( \frac{y}{%
2T}\right) }\right] \mbox{Re}\frac{\tilde{\omega}_{j}(y)}{\sqrt{\tilde{\omega%
}_{j}^{2}(y)-\Phi _{j}^{2}(y)}},  \label{ImEl}
\end{eqnarray}%
where ${\Phi}_{i}(\omega )$ and $\tilde{\omega}_{i}(\omega )\equiv
Z_{i}(\omega )\omega $ are the renormalized gap function and the
renormalized frequency respectively, $\Gamma _{ij}$ denotes the
impurity scattering rate within the Born approximation. The real
and the imaginary
parts of the Eliashberg functions ${\Phi}_{i}(\omega )$ and $\tilde{%
\omega}_{i}(\omega )$ are connected by the Kramers-Kronig
relations. Hence, they have the same Fourier images. This yields a
procedure for a rapid solution. The convolution type integrals
(Eqs. \ref{eq1}-\ref{ImEl}) should be calculated by the Fast
Fourier Transform (FFT) algorithm. The inverse \textit{complex}
Fourier transformations of the results obtained give
\textit{complex} values of $\Phi _{i}(\omega )$ and $\tilde{\omega}%
_{i}(\omega )$.

\section{Interband Scattering}

\textit{Intraband} scattering from nonmagnetic impurities does not
affect $T_{\mathrm{c}}$ and the superconducting densities of states
$N_{i}(\omega)$, as well as the thermodynamic potentials. However,
\textit{interband} scattering is expected to modify $T_{\mathrm{c}}$ and
$N_{i}(\omega)$ strongly. In the weak coupling regime this effect has been
demonstrated in Refs.~\onlinecite{Allen,Schopohl,GolMaz}. In the following
we will calculate $T_{\mathrm{c}}$, the gap functions and the superconducting
DOS by solving the nonlinear equations (Eq. \ref{reEEa} - \ref{ImEl}) for
various values of the interband \textit{nonmagnetic} scattering rates
$\Gamma _{\sigma \pi }$ and $\Gamma _{\pi \sigma }$. For convenience, we
define an interband scattering parameter $\Gamma $ in the following way
$\Gamma _{\sigma \pi }=\Gamma N_{\pi }(0)/N_{tot}(0)$, $\Gamma _{\pi \sigma
}=\Gamma N_{\sigma }(0)/N_{tot}(0)$.

\subsection{Gap Functions and the Density of States.}

\begin{figure}[hbtp]
\includegraphics[width=2.8in ]{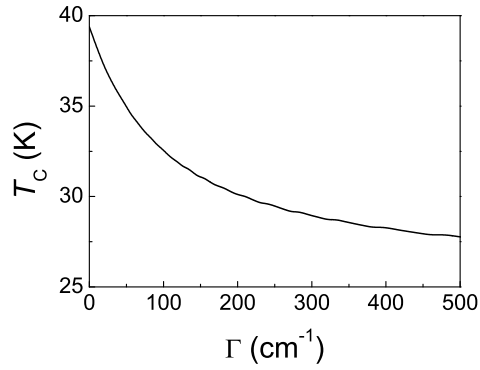}
\caption{The dependence of $T_c$ on the interband scattering parameter $\Gamma$.}
\label{fig:Tc}
\end{figure}

As shown in Fig.~\ref{fig:Tc}, $T_{\mathrm{c}}$ gradually
decreases with increasing $\Gamma $ and saturates at a value
corresponding to that expected for isotropic coupling. The initial
decrease of $T_{\mathrm{c}}$
with $\Gamma$ amounts to
$T_{\mathrm{c}}(\Gamma) - T_{\mathrm{c}}(0)= 0.10(1)\,\mathrm{K/cm}^{-1}\cdot\Gamma$,

\begin{figure}[hbtp]
\includegraphics[width=2.8in]{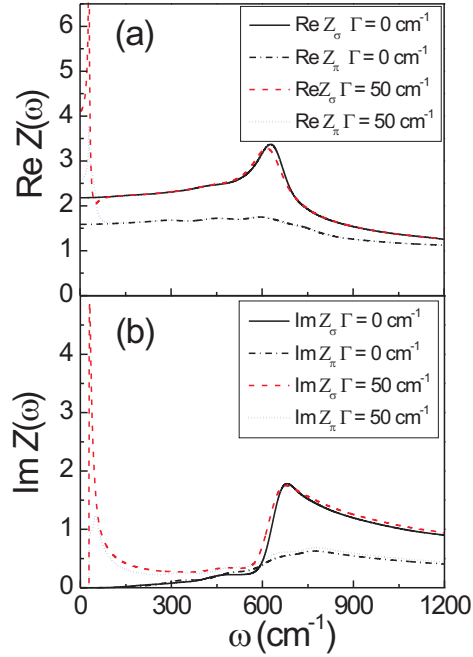}
\caption{The energy dependencies of the renormalization functions in a
two-band model at $T/T_{\mathrm{c}}=0.1$ in the clean limit $\Gamma$\,=\,0
and for $\Gamma$\,=\,50\,cm$^{-1}$.}
\label{fig:Z}
\end{figure}

\begin{figure}[hbtp]
\includegraphics[width=2.8in]{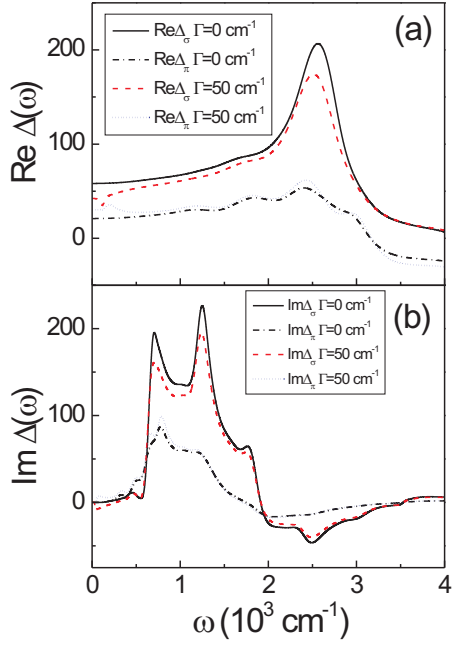}
\caption{The energy dependencies of the gap functions
obtained by using a strong coupling two-band Eliashberg model at
$T/T_{\mathrm{c}}=0.1$ in the clean limit and with
$\Gamma$\,=\,50\,cm$^{-1}$.}
\label{fig:Delta}
\end{figure}

The DOS in the superconducting  state, $N_{i}(\omega )$, is given by the expression
\begin{equation}
N_{i}(\omega )=N_{i}(0)\mbox{Re}\frac{Z_{i}(\omega )\,\omega }{\sqrt{%
Z_{i}^{2}(\omega )\,\omega ^{2}-\Phi _{i}^{2}(\omega )}}  \label{DoS}
\end{equation}%
where $N_{i}(0)$ is the DOS in the normal state at the Fermi level of the
corresponding energy band. Here $\Phi _{i}(\omega )=\Delta _{i}(\omega
)Z_{i}(\omega )$, where $\Delta _{i}(\omega )$ and $Z_{i}(\omega )$ are
complex pair potentials and renormalization functions. Figs.~\ref{fig:Z} and %
\ref{fig:Delta} display the $Z_{\sigma ,\pi }(\omega )$ and $\Delta _{\sigma
,\pi }(\omega )$ as obtained from using spectral functions calculated from
first-principles for the effective two-band model in MgB$_{2}$ \cite{Golubov}.

The results demonstrate the self-energy effects arising due to the
sizeable electron-phonon interaction in MgB$_{2}$. The real parts
$\Delta _{\sigma ,\pi }(\omega )$ and $Z_{\sigma ,\pi}(\omega )$
strongly depend on $\omega$ when $\omega $ becomes comparable to
the characteristic phonon frequencies. The imaginary parts
appearing at these energies indicate the decay of quasi-particles
due to this strong interaction. Furthermore, the effects of
impurity scattering are also visible as additional structure at
low energies comparable to the scattering rate $\Gamma$. This
structure is particularly strong in the real and imaginary parts
of $Z_{i}(\omega )$. The latter can
be seen from the last term in Eq.~\ref{reEEa}. The impurity contribution to
$\mbox{Im}\,Z_i (\omega)$ is proportional to $\Gamma\,N_j(\omega)$,
where $i, j$ belong to different bands.

\begin{figure}[hbtp]
\includegraphics[width=2.8in ]{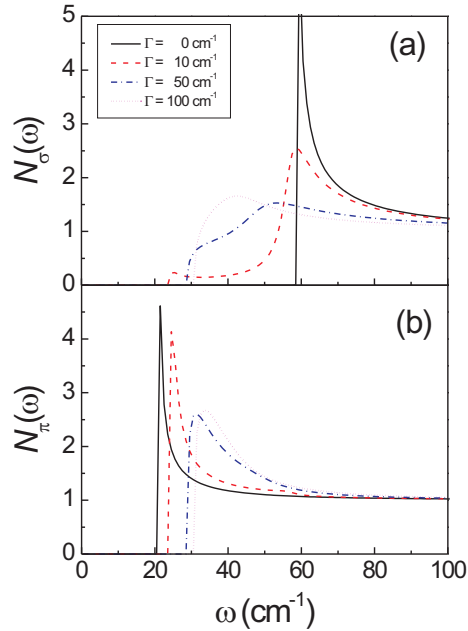}
\caption{The evolution of the low-temperature densities of states with the
interband scattering rate $\Gamma$
in a strong coupling two-band Eliashberg model at
$T/T_{\mathrm{c}}=0.1$.
The upper panel a) shows the superconducting DOS for the
$\sigma$-band, the lower b) the same for the $\pi$-band.}
\label{fig:DoS}
\end{figure}

Fig.~\ref{fig:DoS} shows the  densities of states for different
magnitudes of the interband scattering rate $\Gamma $ at
low-temperature ($T/T_{\mathrm{c}}=0.1$). In the clean limit, the
two bands show two different excitation gaps. In accordance with
earlier calculations \cite{Schopohl,GolMaz}, the interband
impurity scattering mixes the pairs in the two bands, so that the
states appear in the $\sigma$-band at the energy range of the
$\pi$-band gap. These states are gradually filled in with
increasing  scattering rate. At the same time the minimal
$\pi$-band gap in the DOS raises due to increased mixing to the
$\sigma $-band with stronger electron-phonon coupling. Thus the decrease
in $T_{\mathrm{c}}$ is accompanied by an increase of the minimal
gap in the excitation spectrum as has been observed by Gonnelli
\textit{et al.} in Ref. \onlinecite{Gonelli1} and theoretically supported
by some of us in Ref.~\onlinecite{PRL2005}.

\begin{figure}[tbph]
\includegraphics[width=2.8in ]{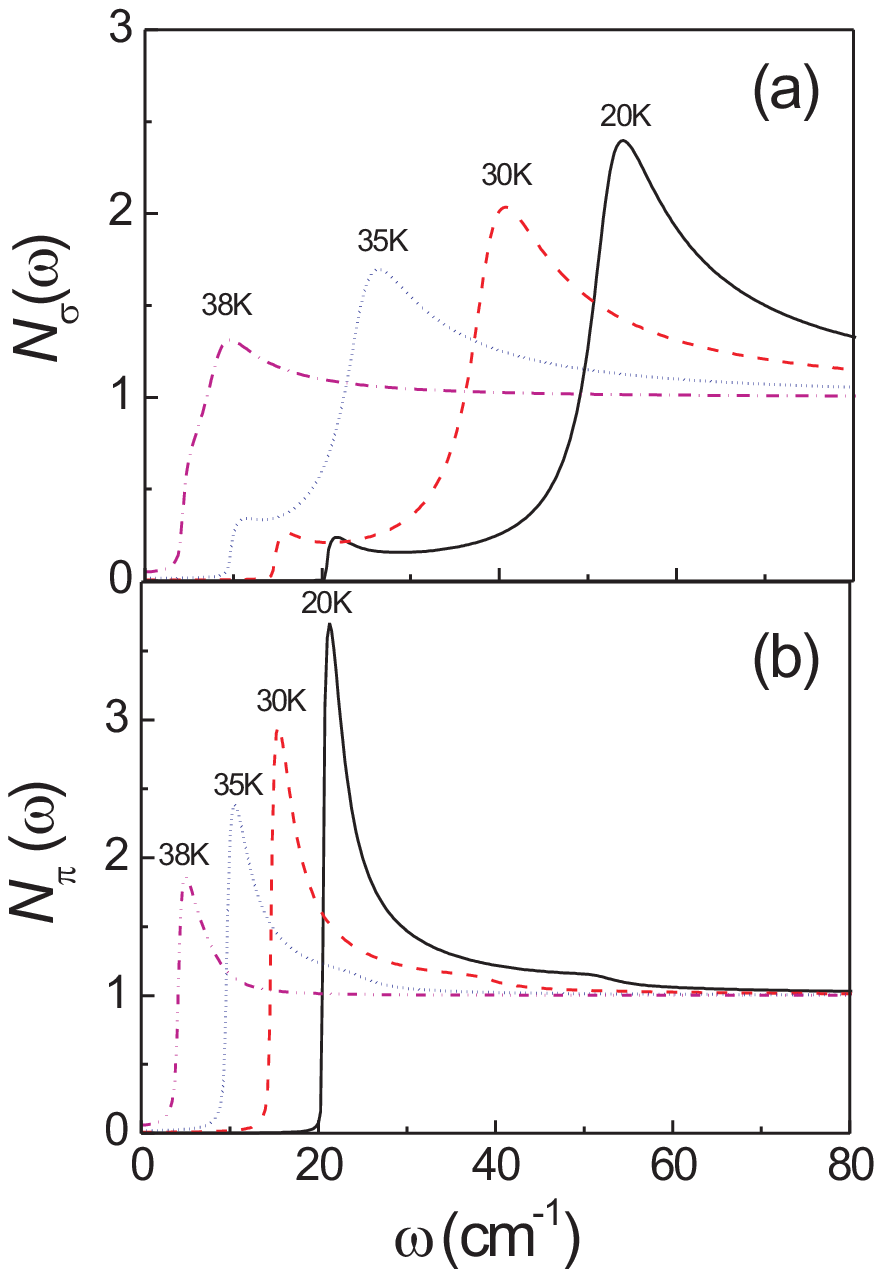}
\caption{The evolution of the superconducting densities of states with
temperature at fixed interband scattering rate $\Gamma $ = 10 cm$^{-1}$
in a strong coupling two-band Eliashberg model.}
\label{fig:DoS_T}
\end{figure}

Fig.~\ref{fig:DoS_T} shows the evolution of the superconducting
DOS with temperature for a fixed values of the interband
scattering rate $\Gamma $ = 10 cm$^{-1}$. One can see that at
finite temperature the densities of states in both bands
become gapless: In addition to the states at the energy range between the
$\pi$-band and the $\sigma$-band gap, states appear down to lowest
energies due to thermal phonons. Such gapless behavior is most
pronounced close to $T_{\mathrm{c}}$. In the isotropic single-band
superconductor, this thermal effect in the strong-coupling regime
was demonstrated earlier in Ref.~\onlinecite{Mikhail}. Note, that the
shape and temperature dependence of the superconducting DOS are
very different compared to the sum of two BCS-like densities of
states. This is particularly pronounced for the $\sigma$-band.
Therefore, one would expect a non-BCS temperature behavior in the
thermodynamical functions.

\subsection{Thermodynamic Functions}

For the numerical calculations for the free energy difference we have used
Eq.~(\ref{eq:wefree}) with parameters described in the previous Section. The
entropy difference between the normal and superconducting states is
determined by the first derivative of the free energy difference with respect
to temperature
\begin{equation*}
\Delta S(T)=d\Delta \Omega (T)/dT,
\end{equation*}%
and the specific heat difference by the second derivative with respect to
temperature%
\begin{equation*}
\Delta C(T)=Td^{2}\Delta \Omega (T)/dT^{2}.
\end{equation*}%
Here we note that taking derivatives from numerically calculated
data (as well as from experimental ones) is often a mathematically
ill-defined or numerically unstable procedure. Therefore, we used
three different schemes to interpolate the numerical data: a) a
Chebyshev scheme to interpolate the free energy calculated at
non-equidistant points $T_j$ = cos($\frac{\pi\,j}{\rm n}$), (j =
0,1, ..., n; where n = $T_c$/$\Delta T$ is the number of points)
and constructing a corresponding matrix n$\times$n operator
\cite{Chebyshev}, b) a polynomial approximation which works well
for large temperatures where the densities of states are smooth
functions without square-root singularities, and c) an
interpolation of the free energy differences by a series of Bessel
functions $K_{1}(n\Delta /T)$ similar to the weak-coupling BCS
approximation (see e.g., Ref.~\onlinecite{AGD}) . The latter
captures the superconducting square-root features and works well
at low temperatures. The data presented below were chosen such
that all three approaches gave similar results.

The specific heat jumps $\Delta C(T_{\mathrm{c}})$ at $T=T_{\mathrm{c}}$
were determined separately by the calculation of the coefficient $\beta =T_{%
\mathrm{c}}\Delta C(T_{\mathrm{c}})/2$ in the term $\Delta \Omega
(T\rightarrow T_{\mathrm{c}})=\beta (1-T/T_{\mathrm{c}})^{2}$.

\begin{figure}[tbp]
\includegraphics[width=2.8in ]{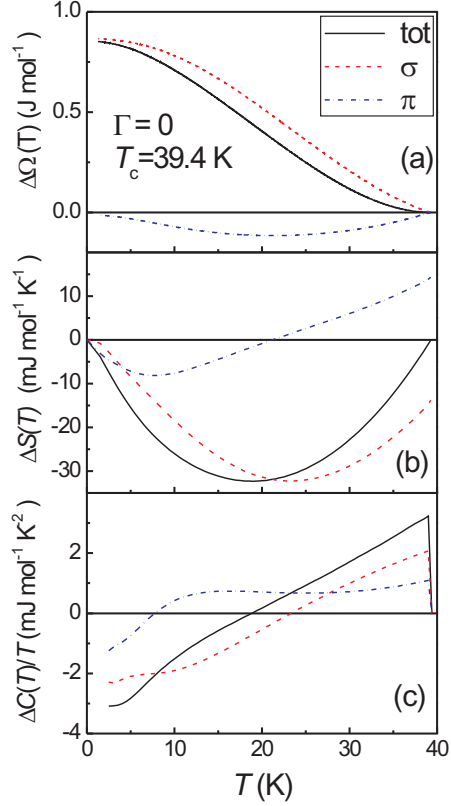}
\caption{Partial and total contributions to the free energy ($\Delta\Omega$%
), the entropy ($\Delta S$) and the heat capacity $\Delta C/T$ in the clean
case.}
\label{fig:partial}
\end{figure}

First, we consider the case without interband impurity scattering ('clean'
case $\Gamma =0$). The expression for $\Delta \Omega (T)$ (see Eq.~\ref%
{eq:wefree}) consists of two terms containing $N_{i}(0)$, the
renormalization factor $Z_{i}$, and the order parameter $\Phi
_{i}$ (or the energy gap $\Delta _{i}=\Phi _{i}/Z_{i}$) for each
band separately. These terms reflect the partial contributions of
each band to the total free energy (cf. upper panel of
Fig.~\ref{fig:partial}). One  sees that the $\pi$-band gives a
negative contribution to the free energy over the full temperature
range. This surprising observation reflects the fact that creation
of the superconducting state in the $\pi$-band and, since the
superconducting state in the $\pi$-band is \textit{induced} by
the occurrence of superconductivity in the strongly interacting $\sigma$%
-band, costs energy.

The coupling due to the nonzero off-diagonal elements of the electron-phonon
interaction is the reason for the same critical temperature $T_{\mathrm{c}}$
and the induced superconducting order parameter in the $\pi$-band. But the
analogy to two-independent contributions to the free energy is not fully
applicable. The partial $\Delta \Omega_{i}$'s near $T_{\mathrm{c}}$ behave as $%
O(T_{\mathrm{c}}-T)$ instead of $\Delta \Omega _{tot}\varpropto O((T_{%
\mathrm{c}}-T)^{2})$, according to the requirements of a mean
field theory. In the middle panel of Fig.~\ref{fig:partial} one
can further see that the entropies $\Delta S_{\sigma ,\pi
}(T_{\mathrm{c}})$ have finite values, whereas only $\Delta
S_{tot}(T_{\mathrm{c}})\equiv 0$, as required by the third law of
thermodynamics. According to this one has to consider only the
total thermodynamic functions as physical ones.

\begin{figure}[tbp]
\includegraphics[width=2.8in ]{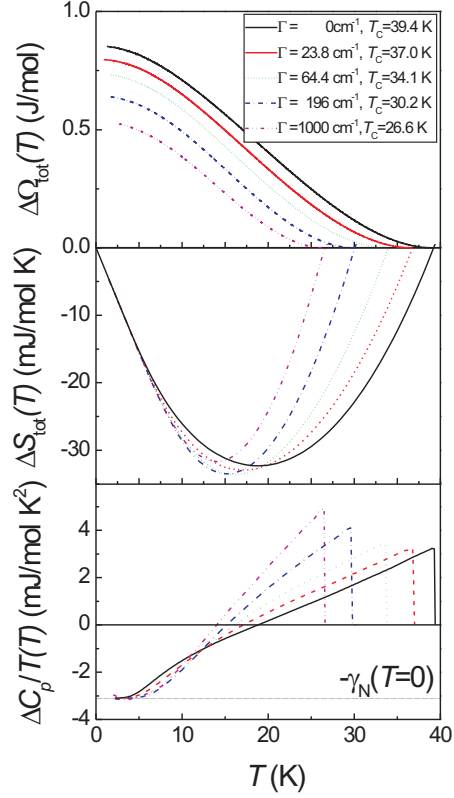}
\caption{Total contributions to the free energy ($\Delta\,\Omega$), the
entropy ($\Delta\,S$) and the heat capacity $\Delta\,C\,/\,T$ for various
impurity scattering rates $\Gamma$. }
\label{fig:freegam}
\end{figure}

Effects of interband impurity scattering on thermodynamic
functions are shown in Fig.~\ref{fig:freegam}. The reduced
specific heat jumps $\Delta C(T_{\mathrm{c}})/T_{\mathrm{c}}$ at
$T=T_{\mathrm{c}}$ grows monotonically with the increase of
$\Gamma $ from $3.24$ mJ mol$^{-1}$K$^{-1}$ in the clean case
\cite{Golubov} to 4.1 mJ mol$^{-1}$K$^{-1}$ for $\Gamma$ =1000
cm$^{-1}$. These values correspond to ratios
$\Delta C(T_{\mathrm{c}})/\gamma (T_{\mathrm{c}})T_{\mathrm{c}}\approx
0.98$ (clean case) and 1.3, which are smaller then the
corresponding BCS value of 1.43 in a single band model. At low
temperatures the ratio $\Delta C(T)/T$ saturates to the value
$\lim_{T\rightarrow 0}(C_{SC}(T)/T-C_{N}(T)/T)=-\gamma
_{N}(T=0)=-\gamma _{0}(1+\lambda _{av})=-3.24$ mJ
mol$^{-1}$K$^{-2}$ which is determined by the bare (band)
electronic specific heat capacity $\gamma _{0}=2\pi
^{2}k_{B}^{2}(N_{\sigma }(0)+N_{\pi }(0))/3,$ and the average
coupling constant
\begin{equation*}
\lambda _{av}=\frac{N_{\sigma }(0)(\lambda _{\sigma \sigma }+\lambda
_{\sigma \pi })+N_{\pi }(0)(\lambda _{\pi \sigma }+\lambda _{\pi \pi })}{%
N_{\sigma }(0)+N_{\pi }(0)},
\end{equation*}%
and does not depend on impurities.

\section{Eliashberg versus Two-Band $\alpha$-model}

Since the two-band $\alpha$-model has been widely used to analyze
the experimental heat capacity data of MgB$_2$ it was interesting
to see to which extend a two-gap $\alpha$-model can reproduce the
Eliashberg results and, if so, how the corresponding parameters
compare with those identified from the Eliashberg calculations.

The $\alpha$-model originally introduced by Padamsee \textit{et
al.}, in close analogy to the BCS theory, assumes a BCS
temperature dependence of the superconducting gap. The magnitude
of superconducting gap at $T$\,=\,0 is
introduced as an adjustable parameter $\alpha$ (from which the $\alpha$%
--model received its name)\cite{Padamsee}. The parameter $\alpha$
is defined according to

\begin{equation}
\Delta(T)\,=\,(\alpha\,/\,\alpha_{\mathrm{BCS}})\,\Delta_{\mathrm{BCS}}
\end{equation}

with $\alpha_{\mathrm{BCS}}$\,=\,$\Delta_{\mathrm{BCS}}(0)$\,/\,$k_{\mathrm{B%
}}T_{\mathrm{c}}$\,=\,1.764 being the weak-coupling value of the gap ratio.

Within the scope of the $\alpha$-model the free energy
$F_{\mathrm{S}}$ in the superconducting state can be written as

\begin{equation}
\Omega_{\mathrm{S}}(T)\,=\,2\,N(0)\,\int\limits_0^\infty\,d\epsilon\,\left\{2%
\,k_{\mathrm{B}}T\,\ln[1\,-\,f(E)]-\frac{1}{2}\,\frac{(E\,-\,\epsilon)^2}{E}%
\,-\,\frac{\Delta^2\,f(E)}{E}\,\right\}  \label{energy}
\end{equation}

with $E$\thinspace =\thinspace $\sqrt{\epsilon ^{2}\,+\,\Delta ^{2}}$ and $%
N(0)$ being the electron and phonon renormalized band-structure electronic
density of states at the Fermi energy.

We subtract the normal state contribution to the free energy which
corresponds to $\Delta$\,=\,0 and introduce the dimensionless parameters $t$%
\,=\,$T\,/T_{\mathrm{c}}$, $\delta(t)\,=\,\Delta(T)\,=\,\Delta_{\mathrm{0}}$
and $x$\,=\,$\epsilon\,/\,k_{\mathrm{B}}T_{\mathrm{c}}$ and arrive at

\begin{equation}
\Delta\,\Omega(t)\,=\,2\,N(0)\,(k_{\mathrm{B}}T_{\mathrm{c}})^2
\int\limits_0^\infty\,dx\,\left\{\,2\,t\,\ln[\frac{1\,-\,f(y,\alpha)}{%
1\,-\,f(x)}]-\frac{1}{2}\,\frac{(y\,-\,x)^2}{y}\,-\, \frac{%
\alpha^2\,\delta(t)^2\,f(y,\alpha)}{y}\,\right\}  \label{differ}
\end{equation}

where $y\,=\,\sqrt{x^{2}\,+\,\alpha ^{2}\delta (t)^{2}}$, $f(y,\alpha
)\,=\,1/[1\,+\,$exp$(y/t)]$, and $f(x)\,=\,f(y,\alpha \,=\,0)\,=\,1/[1+$exp$%
(x/t)]$

The electronic entropy in the superconducting state is obtained from the
first derivative of the free energy with respect to temperature and can be
written as

\begin{equation}
S_{\mathrm{el}}(t)\,/\,\gamma \,T_{\mathrm{c}}=-(3/\pi
^{2})\,\int\limits_{0}^{\infty }\,dx\left\{ f(y,\alpha )\text{ln}%
\,f(y,\alpha )\,+\,(1\,-\,f(y,\alpha ))\text{\thinspace ln\thinspace }%
(1\,-\,f(y,\alpha ))\right\}  \label{entropy}
\end{equation}

wherein the normal-state electronic specific heat capacity
('Sommerfeld\,-\,term') is given by $\gamma$=$\frac{2}{3}\,N(0)\,\pi^2\,k_{%
\mathrm{B}}$.

The electronic heat capacity $C_{\mathrm{el}}$ is calculated from Eq. (\ref%
{entropy}) according to

\begin{equation}
C_{\mathrm{el}}\,/\gamma\,T_{\mathrm{c}}\,=\,t\,(d/dt)\,S_{\mathrm{el}%
}\,/\,\gamma\,T_{\mathrm{c}}  \label{heat}
\end{equation}

To compare the Eliashberg results with the two-gap $\alpha$-model
approximation we have developed least-squares refinement codes to
fit the
entropy (Eq. \ref{entropy}) and the heat capacity (Eq. \ref{heat}) with an $%
\alpha$-model which linearly superposes the contributions from the
$\sigma$ and the $\pi$ electronic system.
For the temperature dependence of the reduced gap $\delta(T)$\,=\,$\Delta(T)$%
\,/\,$\Delta(0)$ we adopted the tabulated values provided by
M\"uhlschlegel \cite{Muhlschlegel}. For the analytical
calculations we used a polynomial fit of the these data.

\begin{figure}[hbtp]
\includegraphics[width=2.8in ]{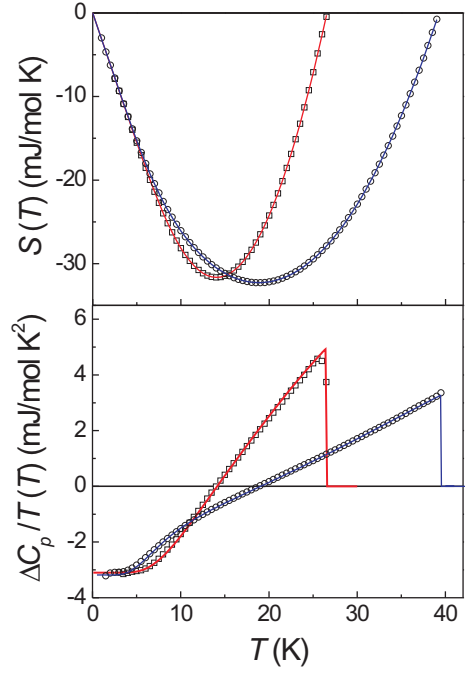}
\caption{(solid lines) Fits of the Eliashberg entropies and the
heat capacities for $\Gamma$\,=\,0 ($T_c$=39.4\,K) and
$\Gamma$\,=\,1000\,cm$^{-1}$ ($T_c$=26.5 K) with the
$\alpha$-model (Eq. \ref{entropy} and Eq. \ref{heat}. For both
entropy data sets ($\Gamma$ = 0 cm$^{-1}$ ($\circ$) and 1000
cm$^{-1}$ ({\tiny{$\square$}}) and the heat capacity data set with
$\Gamma$=0 cm$^{-1}$ ($\circ$) a fit with the two-band
$\alpha$-model was carried out. The fit of the heat capacity data
for $\Gamma$= 1000 cm$^{-1}$ ({\tiny{$\square$}}) was performed
with a single-band $\alpha$-model alone.} \label{SCp}
\end{figure}

The heat capacity was calculated from Eq.\,\ref{heat} using an
appropriate numerical difference quotient as approximation for the
derivative with respect to $t$. Integrations in Eq.(\ref{entropy})
were performed
numerically with a Gaussian quadrature scheme with a cut-off for $x$\,$\geq$%
\,100. Examples for the fits of the entropy and the heat capacity for $%
\Gamma $\,=\,0 and $\Gamma$\,=\,1000\,cm$^{-1}$ corresponding to $T_{\mathrm{%
c}}$\,=\,39.4\,K and $T_{\mathrm{c}}$\,=\,26.5\,K, respectively, are
displayed in Fig.\,\ref{SCp}.

\begin{figure}[tbph]
\includegraphics[width=2.8in ]{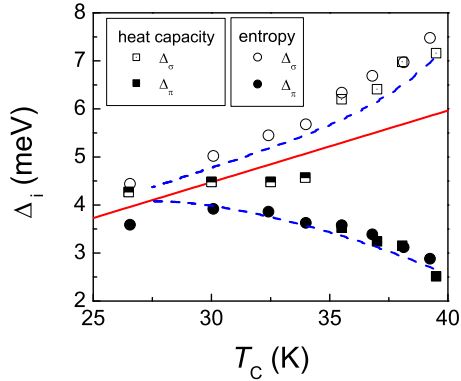}
\caption{Superconducting gaps $\Delta (0)_{\protect\sigma }$ and $\Delta
(0)_{\protect\pi }$ as obtained from the fits of the Eliashberg entropies
($\circ$) and the heat capacities ({\tiny{$\Box $}}) with the
two-band $\alpha$-model (Eq. \ref{entropy} and \ref{heat}). Gap
data gained from fits of the heat capacities for $T_c < 35 K$ were
obtained from a fit with a single one-band $\alpha$-model. The
(red) solid line shows the weak coupling result
$\Delta_{\mathrm{BCS}}$ = 1.76 $T_c$. The
dashed (blue) lines represents the results of the
Matsubara calculations of $\Delta _{\protect\sigma }(\protect\omega _{n}=%
\protect\pi T)$ and $\Delta _{\protect\pi }(\protect\omega _{n}=\protect\pi %
T)$.}
\label{Gaps}
\end{figure}

To fit the data we varied  $\alpha_i$ viz. the energy
gaps $\Delta_i(T)$ and the Sommerfeld constants $\gamma_i$ ($i\,=\,\sigma, \pi$%
) and a single critical temperature $T_c$. All results are
compiled in Table\,\ref{fitres}. Fig.\,\ref{Gaps} displays the fitted gaps
versus $T_c$ viz. the interband scattering parameter $%
\Gamma$.

Attempts to fit also the total energy were less successful and
provided results inconsistent with the results of the entropy and
heat capacity fits. In these fits we observed a tendency to
converge to essentially a single-band model with an averaged gap
somewhat above the weak
coupling BCS result of $\Delta(0)\,/\,k_B T_c$ = 1.76.
Calculations of the free energies with the $\alpha_i$ and
$\gamma_i$ parameters obtained from the fits to the entropies and
heat capacities reproduced the Eliashberg free energies equally well
as obtained from the fits of Eq.\,\ref{energy}.
A closer inspection revealed that  characteristic
differences for various $\Gamma$ are only visible at small
temperatures ($< T_c$/5) where the free energy levels to
saturation. The fits apparently are not sensitive enough to catch
these slight deviations at a satisfying level.

Table\,\ref{fitres} compiles the parameters obtained from the fits
of the entropies and the heat capacities. The results are largely
independent of whether they are obtained from fits of the entropy
or the heat capacity. In general, gaps obtained by fitting of the
entropy are closer to the Eliashberg (Matsubara) gaps.
Fits in the clean case ($\Gamma$\,$\rightarrow$\,0) readily converged
with the parameters listed in Table \ref{fitres}.
For large values of $\Gamma$, convergence of the fits of the
heat capacities with the two-band
model were less stable and fits with a single-band model in some
cases proved to be more conclusive.

\begin{table}[hbtp]
\begin{ruledtabular}
\begin{tabular}{r|ccccccc}
$T_{\mathrm{c}}$ (K) & $\Gamma$ (cm$^{-1}$) & source & $\Delta^{Eliash}_{\sigma}$ (meV) & $%
\Delta_{\sigma}$ (meV) & $\Delta^{Eliash}_{\pi}$ (meV) & $\Delta_{\pi}$ (meV) & $%
\gamma_{\sigma}$/\,$\gamma_{\pi}$ \\ \hline
&  & S &  & 7.48 &  & 2.88 & 0.72\\
\rb{39.4} & \rb{0}  & C & \rb{7.04} & 7.16 & \rb{2.67} & 2.51 & 0.74\\
&  & S & & 6.97 & & 3.12 & 0.63\\
\rb{38.1} & \rb{10}  & C & \rb{6.51} & 7.00 & \rb{2.92} & 3.15 & 0.62\\
&  & S &  & 6.71 &  & 3.40 & 0.64\\
\rb{36.9}&\rb{23.8}  & C & \rb{6.12} & 6.39 & \rb{3.14} & 3.05 & 0.67\\
&  & S &  & 6.34 &  & 3.58 & 0.53\\
\rb{35.5}&\rb{40}  & C & \rb{5.77} & 6.30 & \rb{3.36} & 3.58 & 0.52\\
&  & S &  & 5.68 &  & 3.63 & 0.66\\
\rb{34.0}&\rb{64.4}  & C & \rb{5.47} & 4.57 & \rb{3.56} & - & -\\
&  & S & & 5.45 & 1.38 & & 0.60\\
\rb{32.5}&\rb{100}  & C & \rb{5.95} & 4.51 & \rb{3.57} & - & -\\
&  & S &  & 5.02 & & 3.92 & 0.80\\
\rb{30.1}&\rb{196}  & C & \rb{4.80} & 4.49 & \rb{3.97} & - & -\\
&  & S &  & 4.44 &  & 3.59 & 2.7\\
\rb{26.6}&\rb{1000}  & C & \rb{4.23} & 4.30 & \rb{4.09} & - & -\\
\end{tabular}%
\caption{\label{fitres} Critical temperatures $T_{\mathrm{c}}$,
$\Delta_i$ and Sommerfeld parameters
$\gamma_i$ as obtained from least-squares fit of the
polynom interpolated Eliashberg entropies (S) and the heat
capacities (C) by a two-band $\alpha$--model.
$\Gamma$ is the interband scattering parameter.
The sum $\protect\gamma$\,=\,$\protect\gamma_{\protect\sigma}
+\protect\gamma_{\protect\pi}$ of the fitted Sommerfeld terms
was found to be constant within 3\,\%. For comparison the
Eliashberg (Matsubara) gaps (solutions of Eq. \protect\ref{matsEE}) as displayed in
Fig. \protect\ref{Gaps} are listed.}
\end{ruledtabular}
\end{table}

\begin{figure}[tbp]
\includegraphics[width=2.8in ]{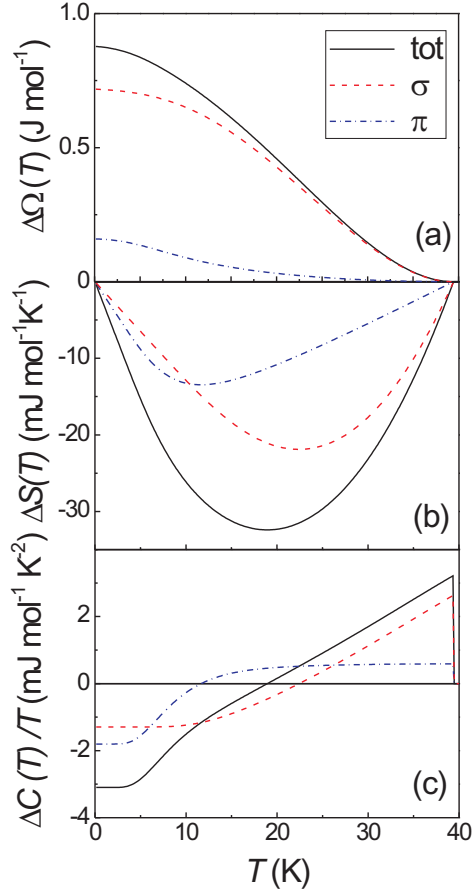}
\caption{Partial and total contributions to the free energy
($\Delta\Omega$), the entropy ($\Delta S$) and the heat capacity
$\Delta C/T$ in the clean case calculated with the two-band $\alpha$-model.}
\label{fig:partcalc}
\end{figure}

In Fig. \ref{fig:partcalc} we show the total and the partial
contributions to the free energy, entropy and the heat capacity
calculated according to the $\alpha$-model using the fitted
parameters given in Table \ref{fitres} for $\Gamma$ = 0 ($T_c$ =
39.4 K). The $\alpha$-model describes the total heat capacity
rather well. There are  subtle differences in the partial $\pi$
and $\sigma$ contributions below $T \approx$ 10 K. These
difference are also reflected in the fitted ratios of the
Sommerfeld constants ($\gamma_{\sigma}$/$\gamma_{\pi}$)$_{\rm
fit}$ which deviates markedly from the ratio of the phonon
renormalized Sommerfeld terms used for the Eliashberg
calculations, ($\gamma_{\sigma}$/$\gamma_{\pi}$)$_{\rm }$ =
$N_\sigma(0)/N_\pi(0)((1 + \lambda_{\sigma\sigma} +
\lambda_{\sigma\pi})/(1 + \lambda_{\pi\pi} + \lambda_{\pi\sigma}))
\approx 1$.

Naturally, since within the scope of the $\alpha$-model all
partial contributions are positive definite, the negative $\pi$
partial free energy and the sign change of the $\pi$ partial
entropy (compare to Fig.\ref{fig:partial}) cannot be reproduced.

Finally, Fig. \ref{Gaps} shows  the superconducting gaps as
obtained from the fits in comparison with the Eliashberg
calculations. The agreement is fairly good for higher $T_c$.
Deviations are seen for $T \cong$ 33 K for the gaps gained from
the fits of the heat capacities, while the gaps received from the
fits of the entropy rather well follow the Matsubara calculations
and  the merging point of both gaps at the weak coupling value is
also well reproduced.

\section{Conclusion}

In summary, using the Eliashberg approach, we have studied the
behavior of the superconducting density of states, energy gaps, free
energy, entropy and specific heat in a strongly-coupled two-band
superconductor with interband impurity scattering. We have
demonstrated strong modifications of the densities of states by
interband scattering and have shown how thermal effects modify
these results. We have calculated the temperature dependencies of
the free energy, the entropy and the specific heat and the
specific jump at $T_c$ as a function of interband scattering rates and
performed a detailed comparison of the phenomenological two-band
$\alpha$-model with the Eliashberg results. We have shown that
despite strong modifications of the DOS by interband scattering,
the $\alpha$-model approach is sufficiently accurate and can -
as a first approximation - be used to extract gap values from
experimental heat capacity or entropy data.

Interband scattering alone, however, is  not sufficient to model
the decrease of $T_c$ observed for Al and C doped  samples
Mg$_{1-x}$Al$_x$(B$_{1-y}$C$_y$)$_2$. As demonstrated recently,
the decrease of $T_c$ can rather be understood in terms of a band
filling effect due to the electron doping by Al and C and a
concomitant scaling of the electron-phonon coupling constant
$\lambda$ by the variation of the density of states as a function
of electron doping \cite{PRL2005}. Compensation of band filling
and interband scattering effects shifts the merging point of the
$\sigma$ and $\pi$ gaps to higher doping concentrations and lower
$T_c$'s than expected based on interband scattering considerations
alone. Only the combination of interband scattering with band
filling effects allowed us to model the nearly constant $\pi$ gap
and the decreased critical temperature and increased doping
concentrations at which the $\sigma$ and $\pi$ gaps finally merge.

\begin{acknowledgments}
We thank I. I. Mazin and M. Putti for stimulating discussions, Y.
Wang for sending us data prior to publication and D. Manske for a
critical reading of the manuscript. Support from the DFG and the
SFB 463 is gratefully acknowledged by S.V. Shulga.
\end{acknowledgments}

\appendix
\section*{Appendix}
In order to calculate the thermodynamic potential $\Omega$ for
a superconductor with strong electron-phonon coupling and nonmagnetic
impurities, we use a general expression for the electron Matsubara
$2\times 2$ matrix Green function
\begin{equation}
\overset{\wedge }{G_{j}}(p)=-\frac{iZ_{jn}^{S}\omega _{n}\hat{\tau}_{0}+\xi
_{j}(\mathbf{p})\hat{\tau}_{3}+\Phi (p)\hat{\tau}_{2}}{(Z_{jn}^{S}\omega
_{n})^{2}+\xi _{j}^{2}(\mathbf{p})+\Phi ^{2}(p)},  \label{G}
\end{equation}%
where $\xi _{j}(\mathbf{p})$ is the bare spectrum  ($p=\{j,\mathbf{p}%
,\omega _{n}\},$ with band index $j$ and momentum $\mathbf{p}$). Pauli matrices $\hat{\tau}$
correspond to Nambu space. This Green function obeys the Eliashberg
equations, which allows us to express the potential $\Omega $ directly
through it as will be shown below.

The thermodynamic potential $\Omega $ can always be expressed by the
electron Green function by means of integration over the electron charge.
Using e.g.\ Eq.(16.9) from Ref.~\onlinecite{AGD} one obtains
\begin{equation}
\begin{array}{c}
\Omega =\Omega _{e}^{(0)}+\Omega _{ph}^{(0)}+T\sum_{p}
\int\limits_{0}^{1}%
\frac{dx}{x}\,\text{tr}\left[ \overset{\wedge }{G}_{(0)}^{-1}(p)\left(
\overset{\wedge }{G}(x,p)-\overset{\wedge }{G}_{(0)}(p)\right) \right] = \\
\Omega _{\text{ }e}^{(0)}+\Omega _{ph}^{(0)}+T\sum_{p}\int\limits_{0}^{1}%
\frac{dx}{x}\,\text{tr}\left[ \overset{\wedge }{\Sigma }(x,p)\overset{\wedge
}{G}(x,p)\right]%
\end{array}
\label{intoverx}
\end{equation}%
where $x$ is a dimensionless factor. $\overset{\wedge }{G}(x,p)$ and
$\overset{\wedge }{\Sigma }(x,p)$ are the exact electron Green function
and the self-energy for the case when the electron charge has
the value $xe$. $\overset{\wedge }{G_{(0)}}$($p$) is the Green function for
zero coupling constant.

The electron-phonon contribution can be expressed in terms of electron and
phonon Green functions by means of Eliashberg equation:
\begin{equation}
\overset{\wedge }{\Sigma }(x,t_{M},\mathbf{r})=T\sum_{\omega _{n},\mathbf{p}%
,j=\pi ,\sigma }\overset{\wedge }{\Sigma }(\omega _{n},\mathbf{p},j)e^{i%
\mathbf{p\cdot r}-i\omega _{n}t_{M}}=x^{2}\hat{\tau}_{3}\overset{\wedge }{G}%
(x,t_{M},\mathbf{r})\hat{\tau}_{3}D(t_{M},\mathbf{r}),  \label{sigmar}
\end{equation}%
where $D(t_{M},\mathbf{r})=g^{2}D_{(0)}(t_{M},\mathbf{r})+D_{imp}(t_{M},%
\mathbf{r})$.
$D_{(0)}(t_{M},\mathbf{r})$ is the phonon Green function expressed in
coordinate representation,$-1/T\leq t_{M}\leq 1/T$ is the Matsubara time.
Here we suppose that the phonon Green function is independent of the
coupling constant in the adiabatic approximation. This is the usual
approximation, which is related to the fact that the electron-phonon
Hamiltonian contains the phonon spectrum already renormalized due
to the electron-phonon interaction and one should not take this renormalization
into account once more. The second term corresponding to impurity
scattering is considered in the Born approximation, where $D_{imp}(t_{M},%
\mathbf{r})\varpropto e^{2}$. Below we follow Ref. \onlinecite{gd}. Making
use of Eq. (\ref{sigmar}) we can derive the simple identity:
\begin{equation*}
\text{tr}\left( \overset{\wedge }{\Sigma }(x)\overset{\wedge }{G}(x)-\frac{x%
}{2}\frac{d\overset{\wedge }{\Sigma }(x)}{dx}\overset{\wedge }{G}(x)\right)
=-\text{tr}\left( \frac{x}{2}\frac{d\overset{\wedge }{G}(x)}{dx}\overset{%
\wedge }{\Sigma }(x)\right) ,
\end{equation*}%
which allows us to rewrite Eq. (\ref{intoverx}) in the coordinate
representation ($r=\{t_{M},\mathbf{r}\}$)
\begin{equation*}
\Omega =\Omega _{\text{ }e}^{(0)}+\Omega _{ph}^{(0)}+\int
dr\int\limits_{0}^{1}\frac{dx}{x}\text{tr}\left( \overset{\wedge }{\Sigma }%
(x)\overset{\wedge }{G}(x)-\frac{x}{2}\frac{d\overset{\wedge }{\Sigma }(x)}{%
dx}\overset{\wedge }{G}(x)+\frac{x}{2}\frac{d\overset{\wedge }{\Sigma }(x)}{%
dx}\overset{\wedge }{G}(x)\right) =
\end{equation*}%
\begin{equation*}
=\Omega _{\text{ }e}^{(0)}+\Omega _{ph}^{(0)}-\int dr\int\limits_{0}^{1}%
\frac{dx}{x}\,\text{tr}\left( \frac{x}{2}\frac{d\overset{\wedge }{G}(x)}{dx}%
\overset{\wedge }{\Sigma }(x)\right) +\int dr\int\limits_{0}^{1}\frac{dx}{x}%
\text{tr}\left( \frac{x}{2}\frac{d\overset{\wedge }{\Sigma }(x)}{dx}\overset{%
\wedge }{G}(x)\right) =
\end{equation*}%
\begin{equation*}
=\Omega _{\text{ }e}^{(0)}+\Omega _{ph}^{(0)}-\frac{1}{2}\int
dr\int\limits_{0}^{1}dx\,\text{tr}\left( \frac{d\overset{\wedge }{G}^{-1}(x)%
}{dx}\overset{\wedge }{G}(x)\right) -\frac{1}{2}\int
dr\int\limits_{0}^{1}dx\,\text{tr}\left( \frac{d\overset{\wedge }{G}(x)}{dx}%
\left( \overset{\wedge }{G}_{(0)}^{-1}-\overset{\wedge }{G}^{-1}(x)\right)
\right)
\end{equation*}%
we can now perform the integration over $x$ exactly and find after
Fourier transformation the required expression:
\begin{equation}
\Omega =\Omega _{e}^{(0)}+\Omega _{ph}^{(0)}+T\sum_{p}\left( \ln \det
\overset{\wedge }{G}\right) -\int dr\left( \ln \det \overset{\wedge }{G}%
_{(0)}\right) -\frac{T}{2}\sum_{p}\text{tr}\left( \overset{\wedge }{G}%
_{(0)}^{-1}\overset{\wedge }{G}-\overset{\wedge }{1}\right) .  \label{Omega}
\end{equation}%
This expression is valid provided that the Green functions $\overset{\wedge }%
{G}$ satisfy the Eliashberg equations.

For the difference of free energies in the S and N-states we have after the
Fourier transformation
\begin{equation}
\Delta \Omega =T\sum_{j,\mathbf{p},\omega _{n}}\left( \ln \frac{\det \overset%
{\wedge }{G}_{j}}{\det \hat{G}^{N}_{j}}\right) -\frac{T}{2}\sum_{j,\mathbf{p}%
,\omega _{n}}\text{tr}\left( \overset{\wedge }{G_{0}}_{(j)}^{-1}\left(
\overset{\wedge }{G_{j}}-\overset{\wedge }{G^{N}_{j}}\right) \right) ,
\label{diff}
\end{equation}%
where $\overset{\wedge }{G^N}$ is given by
\begin{equation*}
\overset{\wedge }{G^N_{j}}=-\frac{iZ_{jn}^{N}\omega _{n}\hat{\tau}_{0}+\xi
_{j}(\mathbf{p})\hat{\tau}_{3}}{(Z_{jn}^{N}\omega _{n})^{2}+\xi _{j}^{2}(%
\mathbf{p})}
\end{equation*}%
Eq.\ref{diff} is the sum of Green functions in different bands.
Finally $\Delta \Omega $ can be expressed as
\begin{eqnarray*}
\Delta \Omega =-T\sum_{\omega _{n}}\sum_{j,\mathbf{p}}\ln \left[ \frac{%
(Z_{jn}^{S}\omega _{n})^{2}+\xi _{j}^{2}(\mathbf{p})+\Phi _{jn}^{2}}{%
(Z_{jn}^{S}\omega _{n})^{2}+\xi _{j}^{2}(\mathbf{p})}\right] +
\\
+ T\sum_{\omega
_{n}}\sum_{j,\mathbf{p}}\left[ \frac{-Z_{jn}^{S}\omega
_{n}^{2}+(Z_{jn}^{S}\omega _{n})^{2}+\Phi _{jn}^{2}}{(Z_{jn}^{S}\omega
_{n})^{2}+\xi _{j}^{2}(\mathbf{p})+\Phi _{jn}^{2}}-\frac{-Z_{jn}^{N}\omega
_{n}^{2}+(Z_{jn}^{N}\omega _{n})^{2}}{(Z_{jn}^{N}\omega _{n})^{2}+\xi
_{j}^{2}(\mathbf{p})}\right] .
\end{eqnarray*}

After integration with respect to the momentum we obtain the expression
\begin{eqnarray*}
\Delta \Omega  &=&-2\pi T\sum_{j,\omega _{n}}^{\left| \omega _{c}\right|
}N_{j}(0)\left[ \sqrt{(Z_{jn}^{S}\omega _{n})^{2}+\Phi _{jn}^{2}}-\left|
Z_{jn}^{N}\omega _{n}\right| \right] + \\
&&+\pi T\sum_{j,\omega _{n}}^{\left| \omega _{c}\right| }N_{j}(0)\left[
\frac{-Z_{jn}^{S}\omega _{n}^{2}+(Z_{jn}^{S}\omega _{n})^{2}+\Phi _{jn}^{2}}{%
\sqrt{(Z_{jn}^{S}\omega _{n})^{2}+\Phi _{jn}^{2}}}-\frac{-Z_{jn}^{N}\omega
_{n}^{2}+(Z_{jn}^{N}\omega _{n})^{2}}{\left| Z_{jn}^{N}\omega _{n}\right| }%
\right]
\end{eqnarray*}%
This expression does not contain any impurities directly.
The effect of \textit{intraband} impurities cancels from Eqs. (3-7).
Also $Z^{S}$, $Z^{N}$, and $\Phi$ do not depend on \textit{intraband} scattering,
however these functions are dependent on \textit{interband} impurity scattering.

The final answer is the expression given in Eq. (1)
\begin{eqnarray*}
\Delta \Omega =-\pi T\sum_{j=\sigma ,\pi }\sum\limits_{n=-\omega
_{c}}^{\omega _{c}}N_{j}(0)\left\{ |Z_{jn}^{N}\omega _{n}|-\left| \omega
_{n}\right| -
\frac{2(Z_{jn}^{S}\omega _{n})^{2}-2\omega _{n}^{2}+2\Phi
_{jn}^{2}}{|\omega _{n}|+\sqrt{(Z_{jn}^{S}\omega _{n})^{2}+\Phi _{jn}^{2}}}+
\right. \\  \left.
+\frac{(Z_{jn}^{S}\omega _{n})^{2}-Z_{jn}^{S}\omega _{n}^{2}+\Phi _{jn}^{2}}{%
\sqrt{(Z_{jn}^{S}\omega _{n})^{2}+\Phi _{jn}^{2}}}\right\} .
\end{eqnarray*}

\end{document}